\documentclass[12pt]{article}
\usepackage{graphicx}
\usepackage{epsfig}
\usepackage[utf8x]{inputenc}
\usepackage[T2A]{fontenc}
\usepackage[english]{babel}
\usepackage{textcomp}
\usepackage{mathtools}
\usepackage{dcolumn}
\usepackage{epsfig}
\usepackage{epstopdf}
\usepackage{bm}
\usepackage[left=0.5cm,right=0.5cm,top=2cm,bottom=2cm]{geometry}
\usepackage[usenames]{color}
\usepackage{colortbl}

\newcommand{\be}{\begin{equation}}
\newcommand{\ee}{\end{equation}}

\newcommand{\bea}{\begin{eqnarray}}
\newcommand{\eea}{\end{eqnarray}}
\newcommand{\beq}{\begin{equation}}
\newcommand{\eeq}{\end{equation}}

\def\fun#1#2{\lower3.6pt\vbox{\baselineskip0pt\lineskip.9pt
		\ialign{$\mathsurround=0pt#1\hfil##\hfil$\crcr#2\crcr\sim\crcr}}}

\begin{document}
\title{Geodesic deviation on symmetry axis in Taub--NUT metric}
	\author{V.P. Vandeev, A.N. Semenova,}
\maketitle
\begin{center}
	{\it Petersburg Nuclear Physics Institute of National Research Centre ``Kurchatov Institute'',\\ Gatchina, 188300, Russia}
\end{center}

\begin{abstract}
An important aspect of General Relativity is to study properties of geodesics. A useful tool for describing geodesic behavior is the geodesic deviation equation. It allows to describe the tidal properties of gravitating objects through the curvature of spacetime. This article focuses on the study of the axially symmetric Taub--NUT metric. We study tidal effects in this metric using the geodesic deviation equation. Radial geodesics along the symmetry axis of spacetime are considered. We show that all spatial components of tidal forces always change sign under the event horizon. We find a solution of the geodesic deviation equation for all geodesic deviation vector components. It allows us to quantify the effect of the NUT-charge on the tidal properties of Taub-NUT metric. And another important feature that we found is the regular behavior of all tidal force components at all points of spacetime.
\end{abstract} 
\section{\label{Int}Introduction}
Within the framework of general relativity many questions about the structure of spacetime are often reduced to solving problems on the behavior of geodesic curves in a certain metric. One way to investigate the properties of geodesics is to consider the geodesic deviation equation, which describes the relative acceleration of two close geodesics. This acceleration is often interpreted synonymously with the tidal one.

Historically, the first static solution to Einstein's equations was the Schwarzschild solution \cite{MS}, which describes the spherically symmetric gravitational field of a massive object. Geodesic deviation equation was considered in Schwarzschild metric in \cite{SSSS}, solutions were found for radial geodesics in terms of elementary functions. It was shown that only radial component of geodesic deviation vector tends to infinity near a physical singularity. Which means unlimited growth of radial tidal force component acting on a freely falling body. Natural generalizations of the Schwarzschild solution were Reissner--Nordstr{\"{o}}m metric \cite{MR}, \cite{MN} and Kottler metric \cite{MKot} (which is more often called Schwarzschild--(Anti)-De Sitter metric). The geo\-desic deviation equation in these metrics is considered in the works \cite{TFrn} and \cite{TFkm} respectively. The article \cite{TFkbh} tells about tidal properties near black hole surrounded quintessence, which is called Kiselev black hole \cite{Kis}.

Tidal effects play a critical role in many astrophysical phenomena. For example, tidal disruption event (or tidal disruption flare) is that the star is destructed by supermassive black hole gravity. Tidal forces acting on a star in the Schwarzschild gravitational field are studied in \cite{TDE}. It should also be noted that the relative tidal acceleration between the two gravitational wave detectors was used to study graviton properties \cite{L1}, \cite{L2}.

It is also worth noting that there are quite interesting works also devoted to the study of geodesic deviation equation in static spherically symmetric spacetimes: Ref.~\cite{TFns} is about tidal force near the naked singularity, Refs.~\cite{TFrbh} and \cite{chh} are about tidal effects in regular black holes, Ref.~\cite{MG} generalized geodesic deviation description of Schwarzschild spacetime with holographic massive gravity, Ref.~\cite{EGB} shows properties of the relative geodesic motion in 4D Einstein--Gauss--Bonnet black hole. In \cite{NN} was considered tidal effects in the vicinity of null naked singularity spacetime. And article \cite{ChTF} is about tidal properties of a black hole surrounded by dark matter halo.

But a particularly important class of static metrics is the spacetimes of axial symmetry. Among them are metric of massive rotating Kerr black hole \cite{Kerr}, vacuum static axial symmetry Taub--NUT spacetime \cite{TNUT}, electrically charged Kerr--Newman metric \cite{KN}. The article \cite{TFaxkerr} contains geodesic deviation equation analysis on the symmetry axis of Kerr spacetime.
The study of tidal effects in axisymmetric spactimes in the vicinity of rotation axis, for example, in the Kerr metric, can make it possible to describe the gravitational principle of astrophysical jets formation. The description of relativistic flow formation in tidal acceleration terms is presented in \cite{J1}, \cite{J2}, \cite{J3}.

It is also should be noted that the geodesic deviation equation finds its application not only in the description of tidal forces, but also in the properties of orbit, which differ little from circular ones. The works \cite{SHef}, \cite{SHefKerr} and \cite{Melk} are devoted to the properties of such orbits, and there the so-called Shirokov effect was described in various metrics.

In our work, the main object of research is the geodesic properties of the Taub--NUT spacetime. This metric is very interesting because, on the one hand, like the Kerr metric, it is an axially symmetric generalization of the Schwazschild metric, and on the other hand, it is somewhat simpler than the Kerr metric, because $g_{tt}$ and $g_{rr}$ metric components are independent on the polar and azimuthal angles, unlike the Kerr metric. We consider the geodesic deviation equation along the symmetry axis of spacetime because this particular case allows us to eliminate the dynamics of geodesics along the azimuthal direction. This paper is organized as follows. In Sec.~\ref{NUT} we demonstrate Taub--NUT metric and its horizons. Sec.~\ref{GE} contains geodesic equations and there we restrict the set of all geodesics to the set of on symmetry axis geodesics without azimuthal dynamics. In Sec.~\ref{TA} we directly discuss the geodesic deviation equation, construct a set of Carter tetrads that allows us to diagonalize the tidal tensor. We also show the areas of tidal compression or extension and regions of growth and decrease of the tidal force components. In Sec.~\ref{SGDE} we solve geodesic deviation equations with respect to the radial variable analytically and numerically to illustrate the behavior of geodesic deviation vector components. Final Sec.~\ref{Conc} ends this paper, there we give an interpretation of the obtained calculations and also indicate potential directions for the development of tidal effects studies.

We use signature of the metric $(+,-,-,-)$ and set Newtonian gravitational constant $G$ and light speed $c$ to $1$ throughout this paper.
\section{\label{NUT}Taub--NUT spacetime}
Taub--NUT metric is an interesting vacuum generalization of Schwarzschild spacetime to the case of axial symmetry. Line element in spherical coordinates is
\begin{equation}\label{NUTmetric}
ds^2=f(r)\bigg[dt+2n\left(1-\cos{\theta}\right)d\phi\bigg]^2-\frac{dr^2}{f(r)}-\left(r^2+n^2\right)\left(d\theta^2+\sin^2\theta d\phi^2\right),
\end{equation}
where
\begin{equation}\label{f}
f(r)=1-\frac{2Mr+2n^2}{r^2+n^2}
\end{equation}
and $n$ is called NUT-charge, $M$ is black hole mass. The horizons are determined by the equation $f(r)=0$ or
\begin{equation}\label{eqhor}
r^2-2Mr-n^2=0.
\end{equation}
Equation roots are
\begin{equation}\label{NUThorizons}
r_{\pm}=M\pm\sqrt{M^2+n^2}.
\end{equation}
But only $r_{+}$ corresponds to the horizon, because $r_{-}<0$ for all values of $n$. To To illustrate the absence of physical singularity at the point $r=0$, one can calculate the curvature invariant Kretschmann scalar
\begin{equation}
\mathcal{K}=R^{\mu\nu\alpha\beta}R_{\mu\nu\alpha\beta}=\frac{48\left(n^2-M^2\right)\left(n^6-15n^4r^2+15n^2r^4-r^6\right)}{\left(r^2+n^2\right)^6}+\frac{192Mn^2r\left(3n^4-10n^2r^2+3r^4\right)}{\left(r^2+n^2\right)^6},
\end{equation}
which is regular at zero at nonzero NUT-charge
\begin{equation}
\mathcal{K}(0)=\frac{48\left(n^2-M^2\right)}{n^2}.
\end{equation}
\section{\label{GE}Geodesic equations}
Using the Hamilton--Jacobi equations, it is easy to obtain the geodesic equations in the Taub--NUT metric
\begin{equation}\label{get}
\frac{dt}{d\tau}=\frac{E}{f(r)}-2n\left(1-\cos{\theta}\right)\frac{L+2nE\left(1-\cos{\theta}\right)}{\left(r^2+n^2\right)\sin^2{\theta}},
\end{equation}
\begin{equation}\label{ger}
\left(\frac{dr}{d\tau}\right)^2=E^2-f(r)\frac{\delta_1 r^2+L^2+K}{\left(r^2+n^2\right)},
\end{equation}
\begin{equation}\label{geth}
\left(\frac{d\theta}{d\tau}\right)^2=\frac{K+L^2-\delta_1 n^2}{\left(r^2+n^2\right)^2}-\frac{\left(L+2nE(1-\cos{\theta})\right)^2}{(r^2+n^2)^2\sin^2{\theta}},
\end{equation}
\begin{equation}\label{geph}
\frac{d\phi}{d\tau}=\frac{L+2nE\left(1-\cos{\theta}\right)}{\left(r^2+n^2\right)\sin^2{\theta}}.
\end{equation}
Where $E$ and $L$ are energy and angular momentum of test particle. $K$ is so-called Carter constant, which is the third integral of motion in axially symmetric spacetime, which was for the first time obtained for the Kerr metric \cite{mtbh}, and parameter $\tau$ is the proper time, i.e. affine parameter along the geodesic curve. Geodesic equations in Taub--NUT spacetime are discussed in detail in \cite{EQNUT}. Parameter $\delta_1$ defines the type of geodesics, for timelike geodesics $\delta_1=1$, for spacelike geodesics $\delta_1=-1$, for lightlike geodesics $\delta_1=0$.

The expressions (\ref{get})-(\ref{geph}) form a unit covariant 4-velocity vector $u^\mu$ tangent to the geodesic.
\subsection*{\label{GEax}Geodesic equations along the symmetry axis of Taub--NUT spacetime}
In this work, we are interested in geodesics along the symmetry axis of spacetime, so we set $L=0$ and $K=\delta_1 n^2$ which allows us to get rid of the dynamics in the azimuthal angle $\theta$ for all point on the axis, i.e. for $\theta=0$ angular velocity component $\dot{\theta}=0$.

So Eqs. (\ref{get})-(\ref{geph}) on the axis of symmetry have the form
\begin{equation}\label{getax}
u^0=\frac{dt}{d\tau}=\frac{E}{f(r)},
\end{equation}
\begin{equation}\label{gerax}
\left(u^1\right)^2=\left(\frac{dr}{d\tau}\right)^2=E^2-\delta_{1}f(r),
\end{equation}
\begin{equation}\label{gethax}
u^2=\frac{d\theta}{d\tau}=0,
\end{equation}
\begin{equation}\label{gephax}
u^3=\frac{d\phi}{d\tau}=\frac{nE}{r^2+n^2}.
\end{equation}
Note that geodesics without angular momentum $L$ still have a nonzero polar component of the angular velocity (\ref{gephax}).
\section{\label{TA}Tidal acceleration}
Before discussing the tidal effects associated with spacetime curvature, it is worth noting that we can define ``Newtoninan acceleration'' $A^r$ as
\begin{equation}
A^r=\ddot{r},
\end{equation}
which can be found from Eq. (\ref{gerax}). It is
\begin{equation}\label{NA}
A^r=\delta_1\frac{-Mr^2+n^2(M-2r)}{(r^2+n^2)^2},
\end{equation}
which gives ``Newtoninan radial acceleration'' for the Schwarz\-schild case $n=0$ and massive test body moving along timelike geodesic with $\delta_1=1$
\begin{equation}
A^r=-\frac{M}{r^2}.
\end{equation}
When we discuss a particle freely falling from a rest state at $r=r_0>r_{+}$, we can indicate the turning point $R^{stop}$ and also show that if it exists then it is unique. Eq.~(\ref{gerax}) at rest point allows us to determine the energy $E^2$ at a distance $r_0$ and then this energy will determine another value of the radius $R^{stop}$ at which the speed will be equal to zero again. Thus it is
\begin{equation}
R^{stop}=\frac{n^2\left(M-r_0\right)}{Mr_0+n^2}.
\end{equation}
It is easy to show that $R^{stop}$ is less than $r_{+}$ for any $r_0$. This means that the speed of a freely falling body can reach zero only under the black hole horizon.

The main object of research in this work is the geodesic deviation equation
\begin{equation}\label{GDE}
\frac{D^2\tilde{\xi}^{\mu}}{d\tau^2}=R^\mu_{.\nu\alpha\beta}u^\nu u^\alpha\tilde{\xi}^\beta,
\end{equation}
where $\frac{D^2}{d\tau^2}$ is covariant derivative along the geodesic, $R^\mu_{.\nu\alpha\beta}$ is Riemann curvature tensor and $u^{\mu}$ is the unit 4-velocity vector tangent to the geodesic defined in (\ref{get})-(\ref{geph}) and in particular along the axis in (\ref{getax})-(\ref{gephax}). $\tilde{\xi}^{\mu}$ is geodesic deviation vector, which describes the distance between points on adjacent geodesics corresponding to the same value of the affine parameter $\tau$.

Eq.~(\ref{GDE}) determines the evolution of the $\tilde{\xi}^\mu$ vector and describes the movement of geodesics relative to each other. Tidal effects in various gravitational fields are described using the geodesic deviation equation in the general theory of relativity. For this reason $R^\mu_{.\nu\alpha\beta}u^\nu u^\alpha$ is called the tidal tensor. To calculate it we present the nonzero components of the Riemann tensor
\begin{subequations}
\begin{equation}\label{nonzerocomp}
R^{0}_{.101}=-\frac{f''}{2f},\;R^{0}_{.202}=R^{1}_{.212}=-\frac{rf'}{2}-\frac{n^2f}{r^2+n^2},\\
\end{equation}
\begin{equation}
R^{0}_{.102}=-\frac{\sin{\theta}}{1+\cos{\theta}}\frac{d}{dr}\left(\frac{n^2f}{r^2+n^2}\right),\\
\end{equation}
\begin{equation}
R^{0}_{.113}=2n\sin^2\frac{\theta}{2}\left[\frac{f''}{f}-\frac{rf'}{f\left(r^2+n^2\right)}-\frac{2n^2}{\left(r^2+n^2\right)^2}\right],
\end{equation}
\begin{equation}
R^{0}_{.123}=\left[2n^2\tan^2\frac{\theta}{2}+\frac{r^2+n^2}{f}\right]\frac{d}{dr}\left(\frac{nf\sin{\theta}}{r^2+n^2}\right),
\end{equation}
\begin{equation}
R^{0}_{.213}=\left[4n^2\tan^2\frac{\theta}{2}+\frac{r^2+n^2}{2f}\right]\frac{d}{dr}\left(\frac{nf\sin{\theta}}{r^2+n^2}\right),
\end{equation}
\begin{equation}
R^{0}_{.223}=2n\sin^2\frac{\theta}{2}\left[2+\frac{8fn^2}{r^2+n^2}+r\left(r^2+n^2\right)\frac{d}{dr}\left(\frac{f}{r^2+n^2}\right)\right],
\end{equation}
\begin{equation}
R^{0}_{.303}=-\frac{rf'\sin^2\theta}{2}+\frac{n^2f}{r^2+n^2}\left(\frac{16fn^2\sin^4\frac{\theta}{2}}{r^2+n^2}+8rf'\sin^4\frac{\theta}{2}-\sin^2\theta\right),
\end{equation}
\begin{equation}
R^{0}_{.312}=\left[2n^2\tan^2\frac{\theta}{2}-\frac{r^2+n^2}{2f}\right]\frac{d}{dr}\left(\frac{nf\sin{\theta}}{r^2+n^2}\right),
\end{equation}
\begin{equation}
R^{1}_{.313}=8n^2ff''\sin^4\frac{\theta}{2}-\sin^2\theta\left[\frac{n^2f}{r^2+n^2}+\frac{rf'}{2}\right],
\end{equation}
\begin{equation}
R^{1}_{.323}=\left(r^2+n^2\right)f\frac{d}{dr}\left(\frac{6n^2f\sin\theta\sin^2\frac{\theta}{2}}{r^2+n^2}\right),
\end{equation}
\begin{equation}
R^{2}_{.323}=\sin^2\theta\left[1-f+\frac{4n^2f}{r^2+n^2}\right]+8n^2f\sin^4\left(\frac{\theta}{2}\right)\frac{rf'(r^2+n^2)+2n^2f}{(r^2+n^2)^2}.
\end{equation}
\end{subequations}
Where $f$ is function defined in Eq.~(\ref{f}), prime means differentiation with respect to $r$. And the curvature tensor indices take values $(0,1,2,3)$, which correspond to a set of coordinates $(t,r,\theta,\phi)$. It is not difficult to see that on the symmetry axis at $\theta=0$ Riemann tensor components are greatly simplified. And only the components from the Eq.~(\ref{nonzerocomp}) remain nonzero on the axis.

\begin{figure}[h!]
\center{\includegraphics[width = 12 cm]{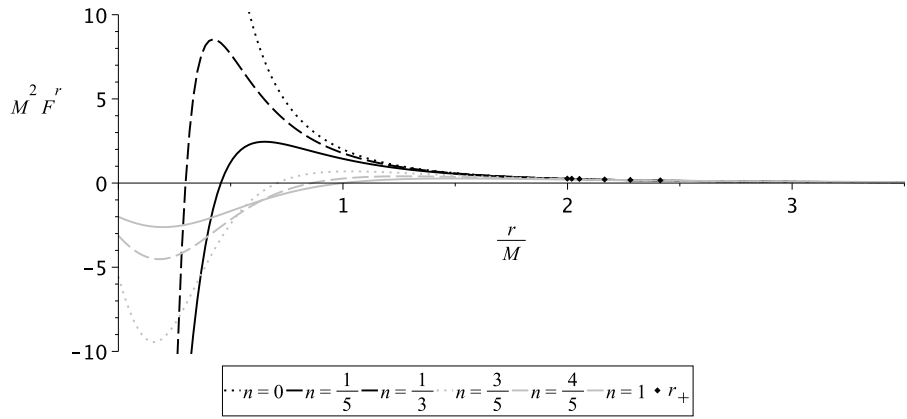}
\caption{\footnotesize{Radial tidal force component for different $n$ per unit length.}}\label{fr}}
\end{figure}

\begin{figure}[h!]
\center{\includegraphics[width = 12 cm]{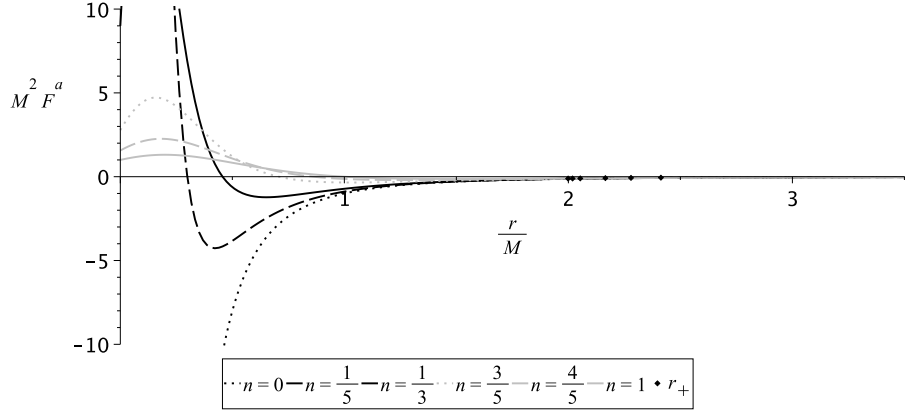}
\caption{\footnotesize{Angular tidal force components for different $n$ per unit length.}}\label{fa}}
\end{figure}

Therefore, the tidal tensor $R^{\mu}_{.\nu\alpha\beta}u^{\nu}u^{\alpha}$ has the form
\begin{equation}\label{TT}
\begin{pmatrix}
\frac{f''\left(E^2-\delta_1 f\right)}{2f}  &-\frac{Ef''\sqrt{E^2-\delta_1 f}}{2f^2}                        &0                                              & 0\\
\frac{Ef''\sqrt{E^2-\delta_1 f}}{2}        &-\frac{E^2f''}{2f}                                             &0                                              & 0\\
0                                          &0                                                              &-\frac{\gamma\delta_1}{2\left(r^2+n^2\right)^2}& 0\\
\frac{\gamma E^2n}{2\left(r^2+n^2\right)^3}&-\frac{\gamma En\sqrt{E^2-\delta_1f}}{2f\left(r^2+n^2\right)^3}&0                                             & -\frac{\gamma\delta_1}{2\left(r^2+n^2\right)^2}
\end{pmatrix},
\end{equation}
where
\begin{equation}
\gamma=r\left(r^2+n^2\right)f'+2n^2f.
\end{equation}
So Eq.~(\ref{GDE}) can be diagonalized using the tetrad basis for free-fall reference frames. Such a basis for an axially symmetric metric is called Carter’s tetrad basis. The procedure for constructing the tetrad basis is described in detail in the appendix of Ref.~\cite{TFaxkerr}. Therefore, we give four tetrad vectors
\begin{subequations}\label{tetrads}
\begin{equation}
e^\mu_t=A\left(r^2+n^2\right)\left(\frac{1}{f},\frac{\sqrt{E^2-\delta_1f}}{E},0,\frac{n}{r^2+n^2}\right),
\end{equation}
\begin{equation}
e^\mu_r=\left(\frac{\sqrt{E^2-\delta_1f}}{f\sqrt{\delta_1}},\frac{E}{\sqrt{\delta_1}},0,0\right),
\end{equation}
\begin{equation}
e^\mu_\theta=\left(0,0,\frac{1}{\sqrt{r^2+n^2}},0\right),
\end{equation}
\begin{equation}
e^\mu_\phi=\left(0,0,0,\frac{1}{\sqrt{\left(r^2+n^2\right)\sin^2\theta-16n^2f\sin^4\frac{\theta}{2}}}\right),
\end{equation}
\end{subequations}
where
\begin{equation}
A=\frac{E}{\sqrt{\delta_1\left(r^2+n^2\right)^2+4E^2n^2\sin^2\frac{\theta}{2}\left[4n^2f\sin^2\frac{\theta}{2}+(r^2+n^2)\left(1+\sin^2\frac{\theta}{2}\right)\right]}}.
\end{equation}
Tetrad set (\ref{tetrads}) satisfy normalization condition $\eta_{\alpha\beta}=e^\mu_\alpha e^\nu_\beta g_{\mu\nu}$ and $\eta_{\alpha\beta}$ is Minkowski metric. All of the above allows us to replace the
geodesic deviation vector $\tilde{\xi}^\mu$ with tetrads from Eqs.~(\ref{tetrads}) as follows
\begin{equation}
\tilde{\xi}^\mu=e^{\mu}_\alpha\xi^\alpha.
\end{equation}
Eq.~(\ref{GDE}) in terms of the vector $\xi^\alpha$ reads
\begin{equation}\label{rgde}
\ddot{\xi}^r=-\frac{\delta_1f''}{2}\xi^r,
\end{equation}
\begin{equation}\label{angde}
\ddot{\xi}^a=-\delta_1\left(\frac{rf'}{2\left(r^2+n^2\right)}+\frac{n^2f}{\left(r^2+n^2\right)^2}\right)\xi^a,
\end{equation}
where $a=\theta,\phi$ correspond to the angular components of the geodesic deviation vector. The dot denotes the derivative with respect to the $\tau$. And the time component of the geodesic deviation vector has trivial equation $\ddot{\xi}^t=0$. Eqs.~(\ref{rgde}) and (\ref{angde}) demonstrate that lightlike geodesics (at $\delta_1=0$) have trivial equations $\ddot{\xi}^\mu=0$ for all components $\mu=t,r,\theta,\phi$. Therefore, the main interest for us is timelike geodesics at $\delta_1=1$. Components of tidal acceleration in the Taub--NUT metric are
\begin{equation}\label{tar}
\ddot{\xi}^r=2\delta_1\left[\frac{Mr^3+n^2\left(3r^2-3Mr-n^2\right)}{(r^2+n^2)^3}\right]\xi^r,
\end{equation}
\begin{equation}\label{taa}
\ddot{\xi}^a=-\delta_1\left[\frac{Mr^3+n^2\left(3r^2-3Mr-n^2\right)}{(r^2+n^2)^3}\right]\xi^a.
\end{equation}
Since for $\delta_1=0$ the right sides of Eqs.~(\ref{tar}) and (\ref{taa}) are trivial, and nonzero values of $\delta_1$ differ in sign, we represent in the Figs.~\ref{fr} and \ref{fa} spatial component of tidal force with $\delta_1=1$. Note also that expression (\ref{tar}) is different from (\ref{NA}).

It is seen that at zero NUT-charge $n=0$ spatial components of tidal force has form
\begin{subequations}
\begin{equation}
\ddot{\xi}^r=\frac{2M\delta_1}{r^3}\xi^r,
\end{equation}
\begin{equation}
\ddot{\xi}^a=-\frac{M\delta_1}{r^3}\xi^a,
\end{equation}
\end{subequations}
which are the same as the classic result from \cite{MTW} for timelike geodesics with $\delta_1=1$. It should be noted that at a nonzero NUT-charge the investigated tidal forces do not become infinite at the point $r=0$. They are
\begin{subequations}
\begin{equation}
\ddot{\xi}^r\bigg|_{r=0}=-\frac{2\delta_1}{n^2}\xi^r,
\end{equation}
\begin{equation}\label{singang}
\ddot{\xi}^a\bigg|_{r=0}=\frac{\delta_1}{n^2}\xi^a.
\end{equation}
\end{subequations}
This is an important difference between the Taub--NUT spacetime and Schwarzschild spacetime, where all tidal force components have non-analytical behavior in $r=0$. However, their common feature is the regular behavior of tidal forces on the event horizon. Another important feature of the Taub--NUT metric is that the presence of the NUT-charge makes the spatial components of tidal forces alternating in sign, in contrast to the Schwarzschild metric, where all components were sign constant.

The points at which the tidal acceleration changes sign are determined by the equation
\begin{equation}\label{TFnull}
Mr^3+3n^2r^2-3Mn^2r-n^4=0
\end{equation}
for all spatial components (\ref{tar}) and (\ref{taa}) of geodesic deviation vector. The roots of this equation define the points at which tidal stretching replaces tidal compression. The solution of this equation with respect to the variable $r$ is rather cumbersome, but we can see that the Eq.(\ref{TFnull}) is biquadratic with respect to the variable $n$. Therefore, we can explicitly express the NUT-charge
\begin{equation}\label{NUTnull}
n^2=\frac{3}{2}\left(r^2-Mr\right)\left\{1+\sqrt{1+\frac{4Mr^3}{9\left(r^2-Mr\right)^2}}\right\},
\end{equation}
where we dropped the root with minus before radical because the radical is greater than one, which makes the square of NUT-charge $n^2$ less than zero. Therefore solution (\ref{NUTnull}) exists and is real at $r\ge M$.

\begin{figure}[h!]
\center{\includegraphics[width = 12 cm]{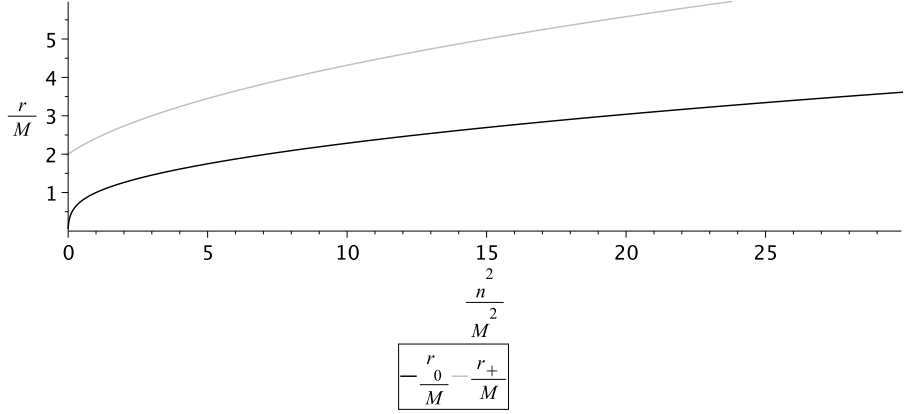}
\caption{\footnotesize{Dependence of the tidal equilibrium radius $r_0$ and the horizon radius $r_+$ on the NUT-charge $n$.}}\label{null}}
\end{figure}

Fig.~\ref{null} shows inverse curve (\ref{NUTnull}) and dependence of horizon radius (\ref{NUThorizons}) on $n^2$. It can be seen that the tidal acceleration zero point is below the event horizon at any values of NUT-charge $n$.

The extremum points of the curves shown in Figs.~\ref{fr} and \ref{fa} are determined by the common equation
\begin{equation}\label{TFextr}
Mr^4+4n^2r^3-6Mn^2r^2-4n^4r+Mn^4=0.
\end{equation}
It is an equation of the fourth degree with respect to $r$, but also biquadratic with respect to $n$. It allows us to express the charge in terms of the radial variable $r$
\begin{equation}\label{NUTextr}
n^2=\frac{2r^3-3Mr^2}{4r-M}\left\{1\pm\sqrt{1+\frac{4Mr-M^2}{\left(2r-3M\right)^2}}\right\}.
\end{equation}

\begin{figure}[h!]
\center{\includegraphics[width = 12 cm]{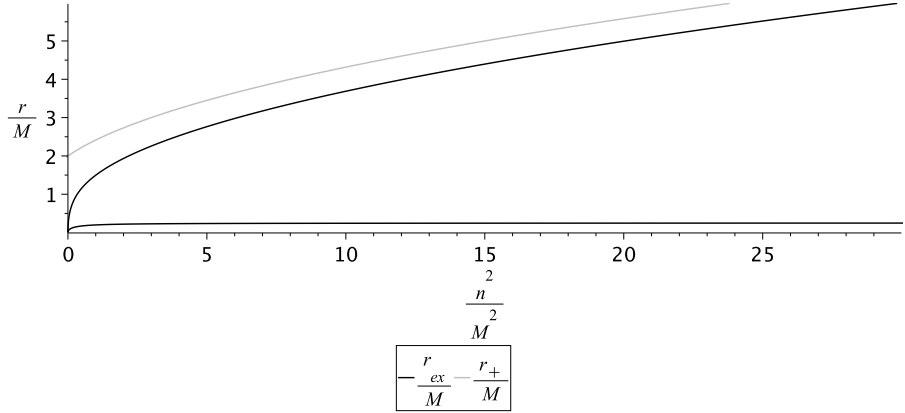}
\caption{\footnotesize{Dependence of the tidal force extremum points $r_{ex}$ and the horizon radius $r_+$ on the NUT-charge $n$.}}\label{extr}}
\end{figure}

Fig.~\ref{extr} shows inverse curve (\ref{NUTextr}) and dependence of horizon radius (\ref{NUThorizons}) on $n^2$. It can be seen that the tidal acceleration extremum points is below the event horizon at any values of NUT-charge $n$.

It is also important to note that formulas (\ref{NUTnull})-(\ref{NUTextr}) and Figs.~\ref{null} and \ref{extr} are valid for both the radial and angular components of tidal forces, because the right-hand sides of expressions (\ref{tar}) and (\ref{taa}) differ only by a factor, which does not affect the position of tidal equilibrium points and extrema of tidal acceleration.
\section{\label{SGDE}Solution of geodesic deviation equation}
In Eqs.~(\ref{rgde}) and (\ref{angde}) below, we pass from differentiation with respect to the affine parameter $\tau$ to differentiation with respect to the radial variable $r$ using Eq.~(\ref{gerax}). So they take the form
\begin{equation}\label{rgde2}
\left(E^2-\delta_1f\right){\xi^r}''-\frac{\delta_1f'}{2}{\xi^r}'+\frac{\delta_1f''}{2}\xi^r=0,
\end{equation}
\begin{equation}\label{angde2}
\left(E^2-\delta_1f\right){\xi^a}''-\frac{\delta_1f'}{2}{\xi^a}'+\left(\frac{\delta_1rf'}{2\left(r^2+n^2\right)}+\frac{\delta_1n^2f}{\left(r^2+n^2\right)^2}\right)\xi^a=0,
\end{equation}
where $a=\theta,\;\phi$. Solutions to these equations will be presented in the following sections.

For the convenience of further calculations, we pass to the dimensionless variables $\rho=\frac{r}{M}$ and $\nu=\frac{n}{M}$. These changes will allow us to eliminate the parameter $M$ from the Eqs.~(\ref{rgde2}) and (\ref{angde2}). Therefore, in all further considerations the prime will mean differentiation with respect to $\rho$ and in new terms function $f$ from (\ref{f}) is
\begin{equation}
f(\rho)=1-\frac{2\rho+2\nu^2}{\rho^2+\nu^2}.
\end{equation}

To compare the new solutions with the previously known ones, we present the solutions of the Eqs.~(\ref{rgde2}) and (\ref{angde2}) without NUT-charge, i.e. in Schwarzschild spacetime for $\nu=0$.
Such solutions are expressed in terms of elementary functions
\begin{equation}\label{elsolr}
\xi^r=A\alpha(\rho)+\frac{B}{\psi^4}\left[6+\psi^2\rho+\frac{3\alpha(\rho)}{\psi}\ln\left(\frac{\alpha(\rho)-\psi}{\alpha(\rho)+\psi}\right)\right],
\end{equation}
\begin{equation}\label{elsola}
\xi^a=\rho\bigg(C+D\alpha(\rho)\bigg),
\end{equation}
where
\begin{equation}
\psi=\sqrt{E^2-1},\;\alpha(\rho)=\sqrt{E^2-1+\frac{2}{\rho}},
\end{equation}
and $A, B, C$ and $D$ are integration constants. These solutions were firstly obtained in Ref. \cite{SSSS}.
In the vicinity of $\rho=0$ they are
\begin{subequations}\label{zerorsing}
\begin{equation}\label{zerorsingr}
\xi^{r}(\rho)=\frac{A\sqrt{2}}{\sqrt{\rho}}+O\left(\sqrt{\rho}\right)\;\text{for}\;\rho\to0,
\end{equation}
\begin{equation}\label{zerorsingaz}
\xi^a(\rho)=\sqrt{2}D\sqrt{\rho}+O\left(\rho\right)\;\text{for}\;\rho\to0.
\end{equation}
\end{subequations}
The same solutions in the vicinity of $\rho=\infty$ behave
\begin{subequations}\label{largersing}
\begin{equation}\label{largersingr}
\xi^{r}(\rho)=\frac{B}{E^2-1}\rho+O\left(1\right)\;\text{for}\;\rho\to\infty,
\end{equation}
\begin{equation}\label{largersingaz}
\xi^a(\rho)=\left(C+D\sqrt{E^2-1}\right)\rho+O\left(1\right)\;\text{for}\;\rho\to\infty.
\end{equation}
\end{subequations}

\subsection{\label{NS} Analytic solutions of the geodesic deviation equation}
\subsubsection{\label{ARC}Radial component}
Note that Eq.~({\ref{rgde2}}) admits an explicit integration
\begin{equation}
\left(E^2-\delta_1f\right){\xi^r}'=\frac{1}{2}\left(E^2-\delta_1f\right)'{\xi^r}+C_1,
\end{equation}
which is a first order inhomogeneous linear differential equation. This type of equations has quadrature solution
\begin{equation}\label{solrgde2}
\xi^r(\rho)=\sqrt{E^2-\delta_1f}\left(C_2+C_1\int\frac{d\rho}{\left(E^2-\delta_1f\right)^{\frac{3}{2}}}\right),
\end{equation}
where $C_1$ and $C_2$ are integration constants. Using the explicit form of the function $f$ from Eq.~(\ref{f}) we will see that the integral in the resulting solution takes the form
\begin{equation}\label{intr}
\int\left(\frac{\rho^2+\nu^2}{(E^2-\delta_1)\rho^2+2\delta_1\rho+(E^2+\delta_1)\nu^2}\right)^{\frac{3}{2}}d\rho.
\end{equation}
This integral is not expressed in terms of elementary functions.

Therefore, we expand the solution (\ref{solrgde2}) in a Taylor series in the vicinity of the points $\rho=0$ and $\rho=\infty$.
\begin{subequations}
\begin{equation}\label{assolrz}
\xi^r=C_2\sqrt{E^2+\delta_1}+\left(\frac{C_1}{E^2+\delta_1}+\frac{C_2\delta_1}{\nu^2\sqrt{E^2+\delta_1}}\right)\rho+O\left(\rho^2\right)\;\text{for}\;\rho\to0,
\end{equation}
\begin{equation}\label{assolrinf}
\xi^r=\frac{C_1}{E^2-\delta_1}\left(\rho-\frac{3\delta_1}{E^2-\delta_1}\ln(\rho)\right)+\left(\frac{\delta_1C_1}{(E^2-\delta_1)^2}+\sqrt{E^2-\delta_1}C_2\right)+O\left(\frac{1}{\rho}\right)\;\text{for}\;\rho\to\infty.
\end{equation}
\end{subequations}
The obtained asymptotics allow us to conclude that there is one more difference between the tidal properties in the Taub--NUT and Schwarzschild metrics. In Schwarzschild spacetime in the vicinity of $r=0$ radial component of geodesic deviation vector behave $\xi^r=\frac{C}{\sqrt{\rho}}+O(\sqrt{\rho})$ \cite{SSSS}, whereas the behavior of this component in the Taub--NUT metric is regularly at zero according to (\ref{assolrz}). But both metrics are asymptotically flat, therefore at spatial infinity the radial component grows linearly in both cases.
\subsubsection{\label{AAC}Angular components}
Since the right-hand side of the angular Eq.~(\ref{taa}) differs from (\ref{tar}) only by a factor, then Eq. (\ref{angde}) can be rewritten as
\begin{equation}\label{angde3}
\ddot{\xi}^a=\frac{\delta_1f''}{4}\xi^a.
\end{equation}
Passing with the help of Eq.~(\ref{gerax}) to differentiation with respect to $r$, in terms of dimensionless variables $\rho$ and $\nu$ we obtain homogeneous second order differential equation
\begin{equation}\label{angeq}
{\xi^a}''+P(\rho){\xi^a}'+Q(\rho)\xi^a=0,
\end{equation}
Where the coefficients $P(\rho)$ and $Q(\rho)$ are functions
\begin{subequations}
\begin{equation}
P(\rho)=-\frac{\delta_1f'}{2\left(E^2-\delta_1f\right)}=\frac{\delta_1\left[-\rho^2+\nu^2\left(1-2\rho\right)\right]}{\left(\rho^2+\nu^2\right)\big[(E^2-\delta_1)\rho^2+2\delta_1\rho+(E^2+\delta_1)\nu^2\big]},
\end{equation}
\begin{equation}
Q(\rho)=-\frac{\delta_1f''}{4\left(E^2-\delta_1f\right)}=\frac{\delta_1\left[\rho^3+\nu^2\left(3\rho^2-3\rho-\nu^2\right)\right]}{\left(\rho^2+\nu^2\right)^2\big[(E^2-\delta_1)\rho^2+2\delta_1\rho+(E^2+\delta_1)\nu^2\big]},
\end{equation}
\end{subequations}
and $a=\theta,\phi$. The angular Eq.~(\ref{angeq}), in contrast to the radial Eq.~(\ref{rgde2}), has no solution in terms of elementary functions even in quadrature form. Therefore, we construct the solution of Eq.~(\ref{angeq}) locally in a vicinity of $\rho=0$ and $\rho=\infty$ using Frobenius method. The essence of the method and the limits of its applicability are succinctly described in the App.~\ref{FEQ}.
\begin{itemize}
\item In the vicinity of the point $\rho=0$ the coefficients $P(\rho)$ and $Q(\rho)$ are expanded in a power series
\begin{subequations}
\begin{equation}
P(\rho)=\frac{\delta_1}{\nu^2\left(E^2+\delta_1\right)}-\frac{2\delta_1\left[\nu^2\left(E^2+\delta_1\right)+\delta_1\right]\rho}{\nu^4\left(E^2+\delta_1\right)^2}+O\left(\rho^2\right),
\end{equation}
\begin{equation}
Q(\rho)=-\frac{\delta_1}{\nu^2\left(E^2+\delta_1\right)}-\frac{\delta_1\left(3E^2+\delta_1\right)\rho}{\nu^4\left(E^2+\delta_1\right)^2}+O\left(\rho^2\right).
\end{equation}
\end{subequations}
The absence of pole terms means that the Frobenius method is applicable. The roots of the equation for powers are $\zeta_1=1$ and $\zeta_2=0$. So form of two linear independent solution are
\begin{subequations}
\begin{equation}\label{flis}
\xi^a_1=\sum_{k=0}^{\infty}c_k\rho^{k+\zeta_1}=\sum_{k=0}^{\infty}c_k\rho^{k+1},
\end{equation}
\begin{equation}\label{slis}
\xi^a_2=\sum_{k=0}^{\infty}d_k\rho^{k+\zeta_2}+A\xi^a_1\ln(\rho)=\sum_{k=0}^{\infty}d_k\rho^k+A\xi^a_1\ln(\rho).
\end{equation}
\end{subequations}
By substituting expression (\ref{flis}) in Eq.~(\ref{angeq}) the coefficients $c_k$ for $k\ge1$ can be expressed in term of $c_0$. Equating them to one we will find the first solution
\begin{equation}\label{fsolz}
\xi^a_1=\rho-\frac{\delta_1\rho^2}{2\nu^2\left(E^2+\delta_1\right)}+\\+\frac{\delta_1\left[\delta_1+\nu^2\left(E^2+\delta_1\right)\right]\rho^3}{2\nu^4\left(E^2+\delta_1\right)^2}+O\left(\rho^4\right).
\end{equation}
The second solution can be found using (\ref{slis}) in Eq.~(\ref{angeq}) assuming $\xi^a_1$ as a solution. Equating the coefficients of all powers of $\rho$ to zero and setting $d_0=1$ and $d_1=0$ (since $d_0$ is free coefficient and $d_1$ is a factor at $r^{\zeta_1}$ in second solution), we get
\begin{equation}\label{ssolz}
\xi^a_2=1+\frac{\delta_1\rho^2}{2\nu^2\left(E^2+\delta_1\right)}+\frac{\delta_1E^2\rho^3}{2\nu^4\left(E^2+\delta_1\right)^2}+O\left(\rho^4\right).
\end{equation}
Therefore, the general solution to Eq.~(\ref{angeq}) is a linear combination of (\ref{fsolz}) and (\ref{ssolz})
\begin{equation}\label{soleqaz}
\xi^a=A_1\xi^a_1+B_1\xi^a_2,
\end{equation}
where $A_1$ and $B_1$ are arbitrary integration constants.
\item To find a solution in the vicinity of spatial infinity change of variables is done $\rho=\frac{1}{s}$. So Eq.~(\ref{angeq}) takes the form
\begin{equation}\label{angeqinf}
{\xi^a}''+W(s){\xi^a}'+H(s)\xi^a=0,
\end{equation}
where the prime denotes differentiation with respect to $s$, coefficients $W(s)$ and $H(s)$ are functions
{\small
\begin{subequations}
\begin{equation}
W(s)=\frac{2}{s}+\frac{\delta_1-\delta_1\nu^2\left(s^2-2s\right)}{\left(1+\nu^2s^2\right)\left[\nu^2s^2\left(E^2+\delta_1\right)+2\delta_1s+E^2-\delta_1\right]},
\end{equation}
\begin{equation}
H(s)=\frac{\delta_1+\delta_1\nu^2\left(3s-3s^2-\nu^2s^3\right)}{s\left(1+\nu^2s^2\right)^2\left[\nu^2s^2\left(E^2+\delta_1\right)+2\delta_1s+E^2-\delta_1\right]}.
\end{equation}
\end{subequations}}
In the vicinity of the point $s=0$ they are expanded into a power series with a pole term
\begin{subequations}
\begin{equation}
W(s)=\frac{2}{s}+\frac{\delta_1}{E^2-\delta_1}+\frac{2\delta_1\left[\nu^2\left(E^2-\delta_1\right)-\delta_1\right]s}{\left(E^2-\delta_1\right)^2}+O\left(s^2\right),
\end{equation}
\begin{equation}
H(s)=\frac{\delta_1}{\left(E^2-\delta_1\right)s}+\frac{\delta_1\left[3\nu^2\left(E^2-\delta_1\right)-2\delta_1\right]}{\left(E^2-\delta_1\right)^2}-\frac{2\delta_1\left[\nu^2\left(3E^4-2E^2\delta_1-\delta_1^2\right)-2\delta_1^2\right]s}{\left(E^2-\delta_1\right)^3}+O\left(s^2\right).
\end{equation}
\end{subequations}
This means that the roots of a quadratic equation determining the leading powers of two linearly independent solutions are $\zeta_1=0$ and $\zeta_2=-1$. Forms of solution are
\begin{subequations}
\begin{equation}\label{flii}
\xi^a_1=\sum_{k=0}^{\infty}c_ks^{k+\zeta_1}=\sum_{k=0}^{\infty}c_ks^{k},
\end{equation}
\begin{equation}\label{slii}
\xi^a_2=\sum_{k=0}^{\infty}d_ks^{k+\zeta_2}+A\xi^a_1\ln(s)=\sum_{k=0}^{\infty}d_ks^{k-1}+A\xi^a_1\ln(s).
\end{equation}
\end{subequations}
By substituting expression (\ref{flii}) in Eq.~(\ref{angeqinf}) the coefficients $c_k$ for $k\ge1$ can be expressed in term of $c_0$ equating which to one we will find the first solution
\begin{equation}
\xi_1^a=1-\frac{\delta_1s}{2\left(E^2-\delta_1\right)}-\frac{\delta_1\left[\nu^2\left(E^2-\delta_1\right)-\delta_1\right]s^2}{2\left(E^2-\delta_1\right)^2}+O\left(s^3\right).
\end{equation}
The second solution can be found using (\ref{slii}) in Eq.~(\ref{angeqinf}) assuming $\xi^a_1$ as a solution. Equating the coefficients at all powers of $\rho$ to zero and setting again $d_0=1$ and $d_1=0$, we get
\begin{equation}
\xi^a_2=\frac{1}{s}-\frac{\delta_1\nu^2s}{2\left(E^2-\delta_1\right)}+\frac{\delta_1E^2\nu^2s^2}{2\left(E^2-\delta_1\right)^2}+O\left(s^3\right).
\end{equation}
In terms of radial variable $\rho$
\begin{subequations}
\begin{equation}\label{fsoli}
\xi_1^a=1-\frac{\delta_1}{2\left(E^2-\delta_1\right)\rho}-\frac{\delta_1\left[\nu^2\left(E^2-\delta_1\right)-\delta_1\right]}{2\left(E^2-\delta_1\right)^2\rho^2}+O\left(\frac{1}{\rho^3}\right),
\end{equation}
\begin{equation}\label{ssoli}
\xi^a_2=\rho-\frac{\delta_1\nu^2}{2\left(E^2-\delta_1\right)\rho}+\frac{\delta_1E^2\nu^2}{2\left(E^2-\delta_1\right)^2\rho^2}+O\left(\frac{1}{\rho^3}\right).
\end{equation}
\end{subequations}
Therefore, the general solution to Eq.~(\ref{angeqinf}) is a linear combination of (\ref{fsoli}) and (\ref{ssoli})
\begin{equation}\label{soleqaninf}
\xi^a=A_2\xi^a_1+B_2\xi^a_2,
\end{equation}
where $A_2$ and $B_2$ are arbitrary integration constants.
\end{itemize}
Summarizing all of the above, we demonstrate the leading order of solutions of Eq.~(\ref{angeqinf})
\begin{subequations}
\begin{equation}\label{spsingsol}
\xi^a=B_1+A_1\rho+\frac{\left(B_1-A_1\right)\delta_1\rho^2}{2\nu^2\left(E^2+\delta_1\right)}+O\left(\rho^3\right)\;\text{for}\;\rho\to0,
\end{equation}
\begin{equation}\label{spinfsol}
\xi^a=A_2+B_2\rho-\frac{\delta_1\left(A_2+B_2\nu^2\right)}{2\left(E^2-\delta_1\right)\rho}+O\left(\frac{1}{\rho^2}\right)\;\text{for}\;\rho\to\infty,
\end{equation}
\end{subequations}
where $A_1,A_2,B_1,B_2$ are integration constants determined by the initial data. Thus, we see that asymptotic flatness of the Taub--NUT spacetime leads to linear increase of the geodesic deviation vector components (\ref{spinfsol}) at spatial infinity. The absence of a physical singularity at the coordinate origin leads to the regular behavior of the the geodesic deviation vector angular components (\ref{spsingsol}). In other words, all angular (\ref{spsingsol}) and radial components (\ref{assolrz}) near zero behave in the same way.

But the given method for solving differential equations is not very illustrative because the solution found by the Frobenius method describes the solution only locally. Therefore, in the next section, we present an alternative approach to the analysis of geodesic deviation.

\subsection{\label{AS} Numerical solutions of the geodesic deviation equation}
To visualize the solutions of the equations and demonstrate the global behavior of the components of the geodesic deviation vectors below, we will solve the Eqs.~(\ref{rgde2}) and (\ref{angde2}) numerically in term of dimensionless variables $\rho$ and $\nu$. Together they can be written as
\begin{equation}\label{fornun}
{\xi^j}''+P(\rho){\xi^j}'+Q_i(\rho)\xi^j=0,
\end{equation}
where $i,j=r,a$ and coefficient $P(\rho)$ is common for all spatial geodesic deviation vector components, but the coefficients $Q_r$ and $Q_a$ are different for radial and angular components
\begin{subequations}
\begin{equation}
P(\rho)=\frac{\delta_1\left[-\rho^2+\nu^2\left(1-2\rho\right)\right]}{\left(\rho^2+\nu^2\right)\big[(E^2-\delta_1)\rho^2+2\delta_1\rho+(E^2+\delta_1)\nu^2\big]},
\end{equation}
\begin{equation}
Q_r(\rho)=\frac{-2\delta_1\left[\rho^3+\nu^2\left(3\rho^2-3\rho-\nu^2\right)\right]}{\left(\rho^2+\nu^2\right)^2\big[(E^2-\delta_1)\rho^2+2\delta_1\rho+(E^2+\delta_1)\nu^2\big]},
\end{equation}
\begin{equation}
Q_a(\rho)=\frac{\delta_1\left[\rho^3+\nu^2\left(3\rho^2-3\rho-\nu^2\right)\right]}{\left(\rho^2+\nu^2\right)^2\big[(E^2-\delta_1)\rho^2+2\delta_1\rho+(E^2+\delta_1)\nu^2\big]},
\end{equation}
\end{subequations}
where $a=\theta,\phi$. Since timelike geodesics are the most interesting from a physical point of view, numerical curves will be plotted at $\delta_1=1$.

\begin{figure}[h!]
\center{\includegraphics[width = 12 cm]{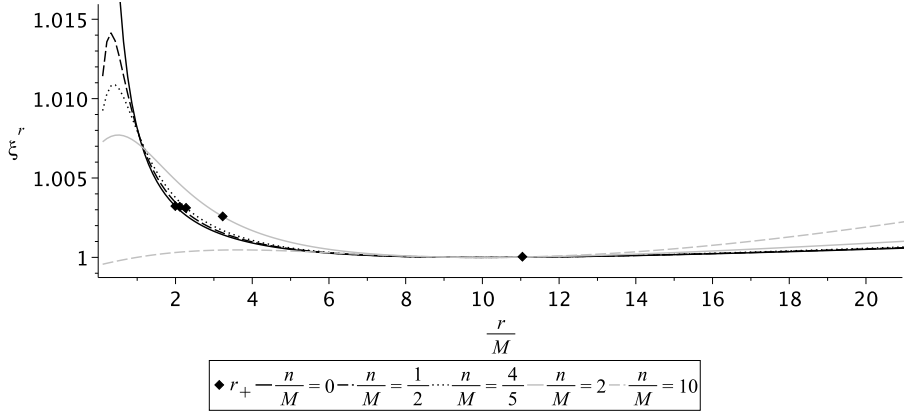}
\caption{\footnotesize{Dependence of $\xi^r$ on $\rho$ at $E=10$. Initial data are $\xi^r(10)=1$, ${\xi^r}'(10)=1$.}}\label{sfr}}
\end{figure}

In Fig.~\ref{sfr} we construct a numerical solution of the radial Eq.~(\ref{fornun}) with $j=r$ for various values of dimensionless NUT-charge $\nu$. It can be seen that the black solid curve with $\nu=0$ in the vicinity of $\rho=0$ corresponds to (\ref{zerorsingr}).

\begin{figure}[h!]
\center{\includegraphics[width = 12 cm]{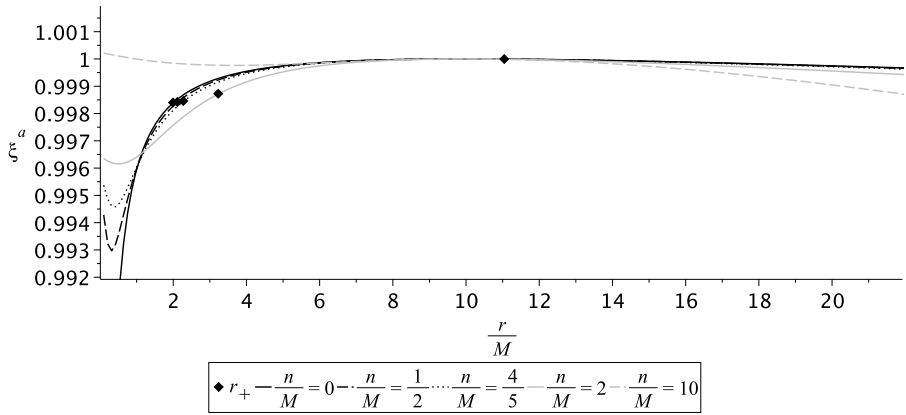}
\caption{\footnotesize{Dependence of $\xi^a$ on $\rho$ at $E=10$. Initial data are $\xi^a(10)=1$, ${\xi^a}'(10)=1$.}}\label{sfa}}
\end{figure}

In Fig.~\ref{sfa} we construct a numerical solution of the angular Eq.~(\ref{fornun}) with $j=a$ for various values of dimensionless NUT-charge $\nu$. It can be seen that the black solid curve with $\nu=0$ in the vicinity of $\rho=0$ corresponds to (\ref{zerorsingaz}).
\section{\label{Conc}Conclusion}
In this paper we investigated tidal effects in Taub--NUT spacetimes along the axis of symmetry. We first demonstrate that the integrals of motion $L$ and $K$ can be chosen so that the geodesic equations (\ref{getax})-(\ref{gephax}) do not have azimuthal dynamics. Further, we show that a freely falling test body in the Taub--NUT metric can stop, but $R_{stop}$ is situated under the horizon. After that, we consider the geodesic deviation equation (\ref{GDE}) on the axis of symmetry of spacetime and the tidal tensor (\ref{TT}) generated by curvature of spacetime. We get Carter’s tetrad basis (\ref{tetrads}), which allows us to diagonalize the geodesic deviation equation. It can be seen that its nontrivial spatial components have form (\ref{tar}) and (\ref{taa}), moreover the polar and azimuthal equations coincide. As in the cases of radial motion in static spherically symmetric spacetime radial and angular geodesic deviation vector components have different signs and are two times different. We show that all components of tidal forces change signs and monotonicity below the event horizon for any NUT-charge value. It is also important to note that in the presence of a nonzero $n$, all components vanish at a single point, but change monotonicity twice.

To demonstrate the behavior of geodesic deviation vector, we solve the geodesic deviation equations (\ref{rgde2}) and (\ref{angde2}) with res\-pect to the radial variable. We present the solutions (\ref{elsolr}) and (\ref{elsola}) known from \cite{SSSS} in the absence of the NUT--charge in order to be able to see how it affects the solution. It turns out that the radial equation (\ref{rgde2}) admits explicit integration, therefore its solution is represented by quadrature (\ref{solrgde2}). Expan\-ding the resulting solution in a Taylor series, we find that in the vicinity of the origin $\xi^r$ behaves regularly (\ref{assolrz}) in contrast to the Schwarzschild and Reissner--Nordstr{\"{o}}m metrics, but similar to the results of Ref.~\cite{TFaxkerr}. And at spatial infinity $\xi^r$ grows linearly (\ref{assolrinf}) because the Taub--NUT metric is asymptotically flat. Angular Eq.~(\ref{angeq}) no longer admits explicit integration. Therefore, we solve it using the Frobenius method in the vicinity of the origin and spatial infinity, because this equation satisfies Fuchs' theorem there. And it turns out that the angular components at zero behave regularly (\ref{soleqaz}) and experience linear growth (\ref{soleqaninf}) at infinity.

Further, we confirm all our analytical calculations by numerical solutions of the Eqs.~(\ref{rgde2}) and (\ref{angde2}). The curves in Figs.~\ref{sfr} and \ref{sfa} fully confirm our analytical conclusions about the singular behavior absence of the geodesic deviation vector in the vicinity of the origin and about the linear growth of all spatial components at infinity.

Despite the large number of articles devoted to the analysis of the geodesic deviation equation and tidal forces in general relativity, there are still many interesting unsolved problems. High-dimensional Tangerlini black holes, black rings, and possibly topological BTZ black holes require the attention of researchers. The behavior of geodesics in spaces with exotic matter is also interesting. Alternatively, one can consider black holes in mimetic gravity. Tidal effects in the vicinity Vaidya dynamic black hole metric are also unexplored.
\section*{Acknowledgement}
We would like to express our gratitude to Yuri Viktorovich Pavlov for meaningful discussions and useful advice.
\appendix
\section{\label{FEQ}Fuchs' equations and Frobenius method}
Second order differential equation for a complex variable function $y(z)$
\begin{equation}\label{FE}
y''(z)+p(z)y'(z)+q(z)y(z)=0
\end{equation}
has a regular singular point at $z=z_0$, if the coefficients $p(z)$, $q(z)$ have pole at $z=z_0$ not higher than the first and second order, respectively. In other words, the coefficients are decomposed into Laurent series as follows
\begin{subequations}
\begin{equation}
p(z)=\sum_{k=-1}^{\infty}a_k(z-z_0)^{k},
\end{equation}
\begin{equation}
q(z)=\sum_{k=-2}^{\infty}b_k(z-z_0)^{k}.
\end{equation}
\end{subequations}
In the vicinity of a regular singular point $z=z_0$ two linearly independent solutions of Eq.~(\ref{FE}) can be found using the generalized series
\begin{subequations}\label{powsol}
\begin{equation}\label{fsff}
y_1(z)=(z-z_0)^{\zeta_1}\sum_{k=0}^{\infty}c_k(z-z_0)^k,
\end{equation}
\begin{equation}
y_2(z)=(z-z_0)^{\zeta_2}\sum_{k=0}^{\infty}d_k(z-z_0)^k,
\end{equation}
\end{subequations}
where $\zeta_1$ and $\zeta_2$ are roots of quadratic equation
\begin{equation}
\zeta(\zeta-1)+a_{-1}\zeta+b_{-2}=0.
\end{equation}
If difference $\zeta_1-\zeta_2$ is not integer, then both solutions of Eq.~(\ref{FE}) presented by power series (\ref{powsol}), but if difference $\zeta_1~-~\zeta_2$ is positive integer, then the first solution for higher $\zeta_1$ remains the same (\ref{fsff}), however the second linearly independent solution has another form
\begin{equation}
y_2=(z-z_0)^{\zeta_2}\sum_{k=0}^{\infty}d_k(z-z_0)^k+Ay_1(z)\ln(z-z_0).
\end{equation}


\begin{thebibliography}{9}
\bibitem{MS}
Schwarzschild K. Berliner Sitzungsbesichte (Phys. Math. Klasse), 189 -- 196 3 Feb. (1916).
\bibitem{SSSS}
Fush H. Ann. Physik, \textbf{495}, 231 -- 233 (1983).
\bibitem{MR}
Reissner H. Ann. Physik, \textbf{50}, 106 -- 120 (1916).
\bibitem{MN}
Nordstr{\"{o}}m G. Proc. Kon. Ned. Akad. Wet., \textbf{20}, 1238 -- 1245 (1918).
\bibitem{MKot}
Kottler F. Ann. Physik, \textbf{56}, 401 -- 462 (1918).
\bibitem{TFrn}
Crispino L.C.B., et al.: Eur. Phys. J. C \textbf{76}, 168 (2016).
\bibitem{TFkm}
Vandeev V.P., Semenova A.N.: Eur. Phys. J. C \textbf{81}: 610 (2021).
\bibitem{TFkbh}
Shahzad M. U., Jawad A., Eur. Phys. J. C \textbf{77}, 372 (2017).
\bibitem{Kis}
Kiselev, V. V.: Class. Quant. Grav. 20, 1187 (2003).
\bibitem{TDE}
Luminet J. P., Marck J .A., Mon. Not. R. Astron. Soc. \textbf{212}, 57 (1985).
\bibitem{L1}
B.P. Abbott, et al., LIGO Scientific, Virgo, Phys. Rev. Lett. \textbf{116} 061102 (2016).
\bibitem{L2}
B.P. Abbott, et al., LIGO Scientific, VIRGO, Phys. Rev. Lett. \textbf{118} 221101 (2017), erratum: Phys. Rev. Lett. \textbf{121} 129901 (2018).
\bibitem{TFns}
A. Goel, R. Maity, P. Roy, and T. Sarkar, Phys. Rev. D \textbf{91}, 104029(2015)
\bibitem{TFrbh}
Sharif M., Sadiq S., JETP \textbf{153}, 232 -- 239 (2018).
\bibitem{chh}
Haroldo C. D. Lima Junior, et al.: Int. J. Mod. Phys. D \textbf{29}, 2041014 (2020)
\bibitem{MG}
Soon-Tae Hong, et al. Phys. Lett. B \textbf{881}  135967 (2020).
\bibitem{EGB}
Jing Li, et al. Eur. Phys. J. C 81, \textbf{590} (2021).
\bibitem{NN}
Siddharth  Madan, Parth  Bambhaniya, https://doi.org/10.48550/arXiv.2201.13163
\bibitem{ChTF}
Jiayi Liu, Songbai Chen, Jiliang Jing,  	
https://doi.org/10.48550/arXiv.2203.14039
\bibitem{Kerr}
Kerr R. P. Phys. Rev. Lett., \textbf{11}, 237 -- 238 September 1. (1963).
\bibitem{TNUT}
E. Newman, L. Tamburino, and T. Unti, J. Math. Phys. \textbf{4}, 915 (1963).
\bibitem{KN}
Newman E. T., Couch E., Chindapared K. et al. J. Math. Phys., \textbf{6}, 918 – 919 (1965).
\bibitem{TFaxkerr}
Haroldo C. D. Lima Junior, et al.: Eur. Phys. J. Plus \textbf{135}: 334 (2020).
\bibitem{J1}
Chicone C., Mashhoon B., Punsly B., Int. J. Mod. Phys. D \textbf{13}, 945 (2004).
\bibitem{J2}
Chicone C., Mashhoon B., Annalen Phys. \textbf{14}, 290 (2005).
\bibitem{J3}
Bini D., Chicone C., Mashhoon B., Phys. Rev. D \textbf{95}, 104029 (2017).
\bibitem{SHef}
Shirokov M. F. Gen. Rel. Grav., \textbf{4}, 131 (1973).
\bibitem{SHefKerr}
Nduka A. Gen. Rel. Grav., \textbf{8}, 347 (1977).
\bibitem{Melk}
Melkumova E.Yu., et al.: Sov. Phys. J., \textbf{33}, 349 (1990).
\bibitem{mtbh}
B. Carter, Phys. Rev. \textbf{174}, 1559 (1968).
\bibitem{EQNUT}
Valeria Kagramanova, et al. Phys. Rev. D \textbf{81}, 124044 (2010).
\bibitem{MTW}
Misner C. W., Thorne  K. S., Wheeler J. A., \textit{Gravitation} (San Francisco: W.H. Freeman, 1973).
\end{thebibliography}
\end{document}